 \documentclass[final,5p,times,twocolumn,authoryear]{elsarticle}

\usepackage{amssymb}
\usepackage{lipsum}

\journal{Icarus}

\usepackage[hidelinks]{hyperref}
\usepackage{amsmath}
\usepackage{comment}

\begin{document}

\begin{frontmatter}



\title{On the rotation of M-type asteroids: statistical evidence for higher rotation rates}


\author[first,fourth]{Fernando Abarzuza\corref{cor1}}
\affiliation[first]{organization={Light Bridges},
            addressline={Observatorio Internacional del Teide}, 
            city={Güímar},
            postcode={38500}, 
            country={Spain}}

\author[second]{Noemí Pinilla-Alonso}
\affiliation[second]{organization={Instituto de Ciencias y Tecnologías Espaciales de Asturias (ICTEA)},
            department={Universidad de Oviedo},
            addressline={Calle de la Independencia, 13}, 
            postcode={33004},
            city={Oviedo},
            country={Spain}}

\author[first,third]{Miquel Serra-Ricart}
\affiliation[third]{organization={Instituto de Astrofísica de Canarias (IAC)},
            addressline={Calle Via Láctea S/N},
            postcode={38205},
            city={La Laguna},
            country={Spain}}

\affiliation[fourth]{organization={Universitat Autònoma de Barcelona (UAB)},
            city={Bellaterra},
            country={Spain}}
\affiliation[fifth]{organization={Departamento de Astrofísica, Universidad de La Laguna (ULL)},
            postcode={38206},
            city={La Laguna},
            country={Spain}}

\author[second]{Santiago Iglesias Álvarez}

\author[first,third,fifth]{Miguel R. Alarcon}

\cortext[cor1]{Currently at IFAE, \url{fabarzuza@ifae.es}}

\begin{abstract}
Rotational dynamics of asteroids carry important information about their internal structure, collisional history and material composition. This work investigates whether M-type asteroids exhibit systematically higher rotation rates than the broader asteroid population. Using Mahlke's probabilistic taxonomy combined with rotation periods and diameters from the Small-Body Database, we analyse the rotational properties of candidate M-type asteroids and compare them with the general asteroid population. Because asteroid rotation strongly depends on size, the comparison is performed using a Monte Carlo resampling approach conditioned on the diameter distribution of the M-type sample. The results indicate that the M-type asteroids rotate, on average, faster than other asteroids of comparable size. While limitations remain due to sample size, heterogeneous data sources, and possible selection effects, the analysis provides statistical evidence that M-type asteroids are associated with higher rotation rates. This finding is consistent with an association between M-type classification and enhanced rotational properties, with potential implications for the internal structure and collisional evolution of small bodies in the Solar System.
\end{abstract}




\end{frontmatter}




\section{Introduction}
\label{introduction}

Asteroids represent the remnant building blocks of the Solar System, 
preserving key information about its early dynamical and thermal evolution. 
Among the diverse taxonomic classes identified in the asteroid belt, M-type asteroids (as defined in this work through the probabilistic taxonomy of \citealp{mahlke2022asteroid}) have long drawn particular attention due to their high radar albedos and spectral properties broadly consistent with metal-rich compositions \citep{ostro1991asteroid,shepard2008radar}. Taxonomic classes, however, are defined spectroscopically rather than compositionally, and the M-class is therefore inherently diverse: several objects classified as M-type are known to exhibit non-metallic or chondritic mineralogies ---a notable example being (21) Lutetia--- reflecting the fact that spectral and albedo criteria are necessary but not sufficient to establish metallic bulk composition. Despite this caveat, the informal designation of M-types as \textit{metallic} asteroids persists widely in the literature, and we sometimes make use of it throughout this work with the same understanding.

Bulk density estimates provide an independent, albeit indirect, line of 
evidence for the partially metal-rich nature of this population. Objects 
belonging to the X complex in the Bus--DeMeo taxonomy \citep{bus2009busdemeo}---broadly encompassing the Mahlke M-type group---tend to exhibit higher average densities than the major silicate and carbon-rich populations, albeit with large dispersion. For instance, \cite{carry2012density} reports mean densities of $2.66 \pm 1.29$~g\,cm$^{-3}$ for S-types and $1.57 \pm 1.38$~g\,cm$^{-3}$ for C-types, compared with $2.94 \pm 0.85$~g\,cm$^{-3}$ for Xe-types and $3.85 \pm 1.27$~g\,cm$^{-3}$ for Xk-types, which are commonly considered the closest Bus--DeMeo analogues to the Mahlke M-type group. Even if these elevated densities are consistent with partial metal enrichment, they remain well below the density of pure iron--nickel material (7--8~g\,cm$^{-3}$), implying a wide variety of compositions and porosities that are not necessarily restricted to purely monolithic metallic bodies.

The study of M-types is therefore of great importance for constraining models of planetary differentiation, the collisional history of the asteroid belt, and the possible existence of remnant metallic cores \citep{rivkin2000nature,matter2013evidence,nichols-fleming2022porosity}. The NASA Psyche mission, currently en route to asteroid (16) Psyche, underscores the relevance of these objects as windows into planetary interiors \citep{bell2021nasa}.

While the mineralogical and geophysical characterisation of M-types has received significant attention, their rotational properties remain less well understood, partly due to the relatively small known population of this class. Rotation rates provide insight into internal strength, collisional evolution, and the role of YORP torques in shaping asteroid spin distributions \citep{Weidenschilling1981fast,carbognani2017asteroids}. Numerous studies have investigated spin barriers and rotation--size trends across asteroid taxonomic classes, yet it remains unclear whether metallic asteroids systematically differ from the broader population. Although faster rotation in M-type asteroids has been suggested in the past \citep{dermott1984asteroid,lagerkvist1998physical}, these results have been inconclusive, largely due to small or biased samples, observational limitations, and the challenges of disentangling size-dependent effects. A hypothetical rotational excess in the M-type population has often been linked to their potentially higher bulk densities and more cohesive or monolithic internal structures compared to S- and C-type asteroids. In principle, such properties would allow metallic bodies to sustain higher rotation rates and approach the cohesionless spin barrier more closely. For typical densities, this limit corresponds to rotation periods of approx. 2.2 h for S-type asteroids \citep{carbognani2017asteroids}. However, observational evidence for such behaviour in the metallic population remains insufficient.

The taxonomy of \citet{mahlke2022asteroid} represents a significant methodological advance over earlier classification schemes: by integrating thousands of spectra and albedo measurements into a unified probabilistic framework, it assigns each asteroid a continuous posterior probability of belonging to each class rather than a discrete label. This enables a controlled treatment of classification uncertainty and maximises the number of candidate M-type asteroids available for statistical analysis. Together with the extensive rotation and diameter data now publicly available, this refined taxonomy enables the re-examination of the long-standing question of whether M-type asteroids rotate faster than the general population using a larger and more representative dataset than was previously possible.

To address these issues, we adopt a statistical approach based on diameter-conditioned resampling of the asteroid population, in which subsets of non-M-type asteroids are repeatedly selected to match the diameter distribution of the M-type sample before comparing their rotation rates. In addition, we perform complementary Monte Carlo experiments without diameter conditioning and a series of robustness tests designed to assess the stability of the result under alternative assumptions, including variation of the classification threshold, an internal catalogue comparison, a leave-one-out jackknife, and a test for the potential influence of asteroid family membership. Together, these analyses aim to provide the most reliable evidence to date on the rotational properties of M-type asteroids.

\section{Data sources}

For this analysis, we combined data from two main sources. The Mahlke catalog, available via VizieR\footnote{\url{https://cdsarc.cds.unistra.fr/viz-bin/cat/J/A+A/665/A26}}, compiles asteroid spectra in the visible and near-infrared together with visual albedo measurements. The authors assembled over 7500 spectra from surveys such as SMASS \citep{xu1995small} and MITHNEOS \citep{binzel2019compositional}, as well as from published datasets and archival observations. After filtering duplicates and low-quality data, the catalogue includes about 5900 spectra for more than 4500 asteroids. Roughly half of these cover only the visible range, while the remainder provide full Vis–NIR coverage. Visual albedos were also obtained from the SsODNet service, based primarily on IRAS, WISE, AKARI, and Spitzer. Spectra and albedos were normalized and merged into a common analytical space, yielding a matrix of 53 spectral dimensions plus one albedo dimension, which served as the basis for latent component analysis and probabilistic classification.

\begin{figure}[h]
	\centering 
	\includegraphics[]{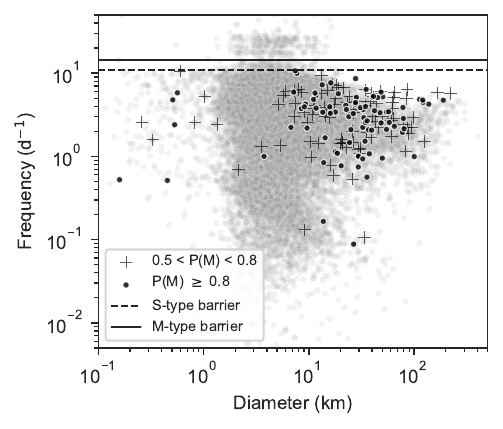}	
	\caption{Rotation frequency versus diameter for the asteroid sample. The background in light grey shows the general asteroid population (19 940 objects from the SBDB with measured diameters and rotation periods). Overplotted are the candidate M-type asteroids according to Mahlke et al. (2022): grey crosses correspond to objects with intermediate probabilities ($0.5 < P(M) \leq 0.8$), while black points mark high-confidence M-types with $P(M) > 0.8$. The dashed black line indicates the cohesionless spin barrier, typical for silicate asteroids, whereas the continuous line indicates a hypothetical limit for M-types, using the X-complex bulk density from \cite{carry2012density}, assuming an idealised spherical body of uniform density, as discussed in Section \ref{sec:robustness}, \textit{Family membership check.}}
	\label{fig:scatter}%
\end{figure}

Complementarily, the Small-Body Database (SBDB) of the \cite{JPL_SBDB} was used to retrieve an updated file of all known small bodies, including rotation periods and diameters when available. The SBDB does not produce new observations itself, but aggregates and reconciles information from multiple sources, such as the Light Curve Database---LCDB, \cite{warner2021lcdb}---and peer-reviewed publications. To ensure the reliability of the rotation periods used in this analysis, only asteroids with an LCDB quality code $U \geq 2$ were retained, excluding poorly constrained or ambiguous period determinations. The resulting dataset is shown in Figure \ref{fig:scatter}, displaying $19\,940$ general asteroids with known diameter and rotation period and 172 M-type asteroids with $P(M)>0.5$---that is, with more than 50\% probability of being M-type---and LCDB quality code higher or equal to 2.  

\section{Statistical analysis}
\label{sec:methods}
Any attempt to assess whether metallic asteroids rotate faster than the general population must confront an immediate challenge: the enormous diversity of diameters in the asteroid population, compared to the evident size bias in the known metallic sample. The rotation frequency of asteroids shows a certain dependence on diameter, with smaller bodies exhibiting, on average, faster spin rates as a consequence of collisional evolution, YORP-driven spin-up, and size-dependent internal structure \citep{pravec2000fast}. A naive comparison between the two groups would therefore risk conflating a genuine compositional signal with a purely size-driven artefact.

This size bias in the metallic sample is a direct consequence of how M-type asteroids are identified. Their classification in \cite{mahlke2022asteroid} relies on visible and near-infrared spectroscopy combined with albedo measurements, techniques that strongly favour large and bright objects accessible to ground-based surveys. As a result, candidate metallic asteroids are heavily concentrated at diameters $D\geq5$ km, as is clearly visible in Figure \ref{fig:scatter} and further analysed in Figure \ref{fig:m-histogram}, while objects below this threshold are virtually absent from the metallic sample---not because small metallic asteroids do not exist, but because they remain largely uncharacterised in terms of spectra and albedo. The general asteroid population is not exempt from observational biases either, but these are less acute and, crucially, not driven by the demanding requirements of spectroscopic taxonomic classification. Recognising and correcting for the size asymmetry between the two samples is therefore a prerequisite for any valid statistical inference.

To define the metallic sample, we adopt a binary classification threshold of $P(M) > 0.5$, the natural Bayesian decision boundary of Mahlke's probabilistic taxonomy. This threshold maximises the number of candidate M-type asteroids available for analysis while ensuring that each included object has a higher posterior probability of being metallic than of belonging to any other class. We acknowledge that this binary treatment discards information contained in the continuous probability distribution; however, the robustness of the results is examined explicitly in Section \ref{sec:robustness} by varying the classification threshold over a wide range of P(M) values; the consistency of the signal across these progressive cuts supports the conclusion that it is not an artefact of borderline classifications.

\subsection{Diameter-conditioned Monte Carlo resampling}
\label{sec:MC}
Monte Carlo resampling is a computational statistical technique in which a large number of random samples are drawn from a reference distribution to build an empirical null, against which an observed statistic can be compared. Rather than relying on analytical approximations of the null distribution—which require assumptions about the underlying data-generating process that may not hold for asteroid populations—resampling methods derive the reference distribution directly from the data, making them robust to non-normality and irregular sample sizes \citep{efron1994introduction}. These properties make them particularly well suited to asteroid population studies, where distributions are typically skewed, sample sizes are modest, and selection effects are difficult to parametrise analytically. Monte Carlo resampling approaches have become a common tool in asteroid rotation dynamics. For instance, \cite{MARZARI2011622} used a Monte Carlo model to simulate the combined effects of YORP and collisions on the rotation rate distribution of Main Belt asteroids, while \cite{durech2022rotation} applied bootstrap resampling to validate rotation periods derived from ATLAS photometry for more than 5000 objects.

\begin{figure}[h]
    \centering
    \includegraphics[scale=1]{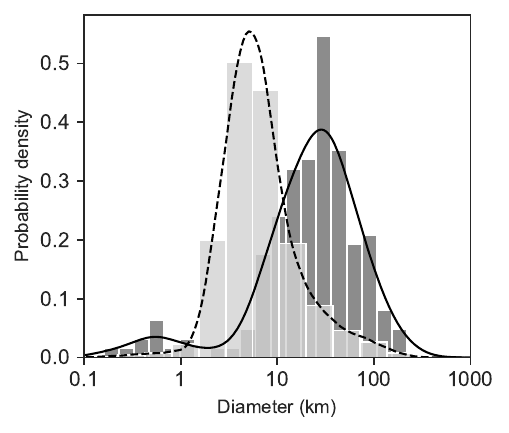}
    \caption{In dark grey, diameter distribution (log-histogram) of the Mahlke population with $P(M)>0.5$ and known diameter and reliable rotation periods (LCDB quality $U\geq 2$). The corresponding kernel density estimate (KDE), black curve, is computed using Scott’s rule \citep{Scott1992}. For comparison, the general asteroid population is shown in light grey, with its KDE overplotted as a dashed curve. M-types are clearly overrepresented at $D>5$ km, reflecting an observational bias in the Mahlke sample. This estimated probability density is used in the Monte Carlo resampling, ensuring a comparison with the metallic sample that is not driven by differences in size distribution.}
    \label{fig:m-histogram}
\end{figure}

In its simplest form, a Monte Carlo comparison between two populations does not account for the size bias described above: if the reference population is drawn freely from the full non-M-type sample, the resulting null distribution would reflect the rotation properties of asteroids across all diameters, rather than those comparable to the known M-type population. To address this, we construct a diameter-conditioned resampling scheme in which non-metallic asteroids are drawn with replacement so that the resulting subsets reproduce the diameter distribution of the M-type sample (Figure \ref{fig:m-histogram}), without relying on discrete binning, as the diameter distribution is estimated via a kernel density estimator---Gaussian kernel, Scott bandwidth, \citep{Scott1992}---from which samples are drawn. Each realization contains the same number of objects as the M-type sample ($n = 172$). In practice, this corresponds to repeatedly drawing 172 asteroids at random with a diameter distribution matched to that of the M-types, which is biased toward larger sizes. By construction, this procedure tests whether the observed rotation properties of the metallic asteroids are compatible with those expected from random samples drawn under the same diameter distribution and sample size.

This process is repeated for 10,000 iterations, each time computing the geometric mean rotation frequency of the resampled non-M-type subset, yielding an empirical null distribution under the hypothesis that composition plays no role once size is held fixed. The geometric mean is adopted as the summary statistic because asteroid 
rotation frequencies follow an approximately log-normal distribution 
\citep[e.g.,][]{pravec2000fast}. For such distributions, the geometric mean is the maximum-likelihood estimator of the location parameter, and it achieves 
minimum variance among the statistics considered here. By contrast, the 
arithmetic mean is a high-variance estimator for log-normal data: a small 
number of extreme draws in any given iteration can substantially inflate 
the sample mean, broadening the null distribution and reducing test 
power. Anyhow, as a robustness check, we repeat the analysis in 
Section \ref{sec:robustness} using the sample median and a 10\,\% 
trimmed mean as alternative statistics.

\begin{figure}[h]
    \centering
    \includegraphics[width=1\linewidth]{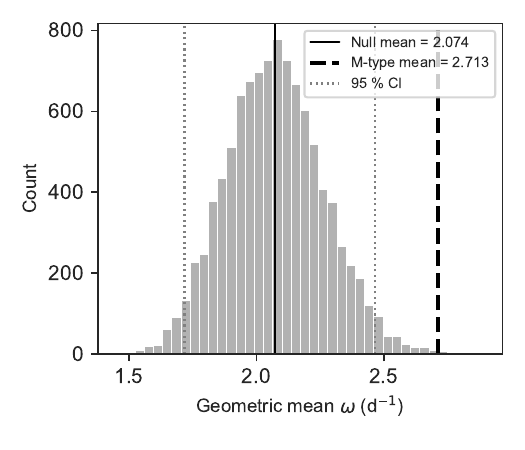}
    \caption{Empirical null distribution of the geometric mean rotation frequency obtained from 10,000 diameter-conditioned Monte Carlo resamplings of the non-metallic population. The shaded histogram shows the distribution of resampled means under the null hypothesis that composition has no effect once size is controlled. The vertical dashed line marks the mean of the M-type sample, while the solid line indicates the mean of the null distribution; the shaded band corresponds to the 95\% confidence interval. The M-type mean lies well beyond the upper tail of the null distribution, indicating a statistically significant excess in rotation rate for M-type asteroids. The corresponding p-value for the M-type mean is 0.0014.}
    \label{fig:MC}
\end{figure}

The results are shown in Figure \ref{fig:MC}. The geometric mean rotation frequency of the M-type sample is $2.713 \;\textrm{d}^{-1}$, while the null distribution has a mean of $2.074 \;\textrm{d}^{-1}$ with a 95\% confidence interval of $[1.718, 2.466] \;\textrm{d}^{-1}$. The metallic mean lies well beyond the upper tail of the null distribution, yielding a one-sided empirical p-value of 0.0014. This result provides statistically significant evidence that M-type asteroids rotate faster than non-M-type asteroids \textit{of comparable size.}

\subsection{Diameter-agnostic Monte Carlo resampling}\label{sec:MC2} As a supplementary consistency check, we perform a second Monte Carlo experiment in which 172 asteroids are drawn at random from the full non-M-type background population with no constraint on their diameter distribution, yielding an unconditioned empirical null distribution. Unlike the conditioned test of Section~\ref{sec:MC}, this approach does not control for the size difference between the M-type sample and the background population, making it intrinsically harder to interpret. It is retained here because it is conservative with respect to the hypothesis under investigation: since M-type asteroids are systematically biased toward larger diameters, and larger asteroids tend to rotate more slowly, the unconditioned null distribution includes a far greater proportion of small, fast-rotating objects, placing the M-type sample at a disadvantage.

\begin{figure}
    \centering\includegraphics{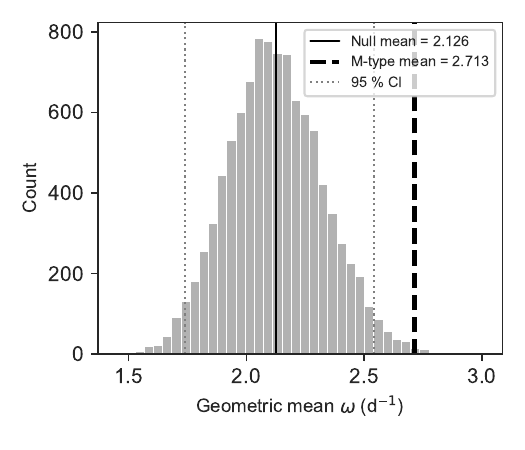}
    \caption{Empirical null distribution of the geometric mean rotation frequency obtained from 10,000 unconditioned Monte Carlo resamplings of the non-M-type population. Unlike the conditioned test (Figure \ref{fig:MC}), reference samples are drawn freely from the full background population with no constraint on diameter, meaning the null distribution reflects asteroids across all size ranges. As expected, this null mean ($2.126\;\textrm{d}^{-1}$) is somewhat higher than its conditioned counterpart ($2.074\;\textrm{d}^{-1}$), reflecting the greater contribution of small, fast-rotating asteroids when no diameter constraint is imposed. The M-type mean still lies well beyond the upper tail of the null distribution, yielding a p-value of 0.0032.}
    \label{fig:MC2}
\end{figure}

The results are shown in Figure \ref{fig:MC2}. The geometric mean rotation frequency of the M-type sample is $2.713\;\mathrm{d}^{-1}$, while the 
unconditioned null distribution has a mean of $2.126\;\mathrm{d}^{-1}$ and 
a 95\% confidence interval of $[1.741, 2.543]\;\mathrm{d}^{-1}$, yielding 
a one-sided p-value of $P = 0.0032$. This result is consistent with that 
of the primary conditioned test ($P = 0.0014$), providing a supplementary 
indication that the rotational excess is not an artefact of the size 
distribution. The unconditioned test compares populations 
with very different diameter distributions and is therefore harder to 
interpret than the diameter-conditioned MC of the previous section. We acknowledge this limitation, but 
emphasize that the resulting bias works against the M-type signal rather 
than in its favour: drawing reference samples freely from the full 
background introduces small, fast-rotating asteroids that raise the null 
mean, so a statistically significant excess under these conditions is a conservative result, even when treated as a supplementary check.

\subsection{Robustness checks}
\label{sec:robustness}
\paragraph{Sensitivity to the choice of summary statistic}

The geometric mean of the rotation frequency was chosen as the test 
statistic because rotation frequencies of small bodies are well described 
by a log-normal distribution \citep[e.g.,][]{pravec2000fast}, for which the geometric mean is the maximum-likelihood estimator of the location 
parameter. To assess sensitivity to this choice, we repeated MC using 
two alternative statistics: the sample median and a 10\,\% trimmed mean 
(i.e., with the lowest and highest 10\,\% of values excluded before 
averaging). Results are summarised in Table~\ref{tab:stats_sensitivity}.

\begin{table}[h]
\centering
\begin{tabular}{lccc}
\hline
Statistic & Observed (d$^{-1}$) & Null (d$^{-1}$) & $p$-value \\
\hline
Geometric mean      & 2.713 & 2.074 & 0.0014 \\
Median              & 3.355 & 2.777 & 0.0168 \\
Trimmed mean & 3.338 & 3.020 & 0.0673 \\
\hline
\end{tabular}
\caption{Sensitivity of the result to alternative summary statistics 
under the diameter-conditioned Monte Carlo test (MC). For each 
statistic, we report the observed value for the M-type sample, the mean 
of the null distribution, and the one-tailed $p$-value. The geometric 
mean is the pre-specified statistic, motivated by the log-normal 
distribution of rotation frequencies; the median and trimmed mean are 
included as robustness checks.}
\label{tab:stats_sensitivity}
\end{table}

The observed value of all three statistics exceeds the null mean, 
confirming that the directional excess is not an artefact of the chosen 
statistic. The geometric mean yields $P = 0.0014$; the median yields 
$P = 0.017$; the trimmed mean yields $P = 0.067$. The geometric mean, being the
natural location parameter of the log-normal, achieves the tightest null
distribution and therefore the highest statistical power. The variation in
p-values reflects differences in power rather than inconsistency in the
underlying signal.
\paragraph{Variation of $P(M)$} A key source of uncertainty in this analysis is the choice of classification threshold $P(M) > 0.5$. To assess whether the results depend on this choice, we repeated the MC test across a wide range of thresholds, from the most restrictive cut ($P(M) > 0.8$, $n = 76$) down to $P(M) > 0.2$ ($n = 276$), as well as the most inclusive possible cut within the Mahlke catalogue ($P(M) > 0$, $n = 1855$), which retains all objects assigned any non-zero M-type probability. The results are summarised in Table \ref{tab:purity}.

Two trends emerge. First, the geometric mean rotation frequency of the sample decreases monotonically as the threshold is lowered, from $2.76\;\mathrm{d}^{-1}$ at $P(M) > 0.6$ to $2.60\;\mathrm{d}^{-1}$ at $P(M) > 0.2$, consistent with progressive contamination by non-metallic objects diluting the signal. Second, and perhaps counterintuitively, the statistical significance of the result improves markedly over this same range: the p-value falls from a marginal p-value of $0.044$ at $P(M) > 0.8$, mostly due to the limited sample size ($n = 76$), to $P < 0.001$ at $P(M) > 0.3$ ($n = 240$). Although the geometric mean of the rotation slowly decreases, the statistical significance is increased by the much larger samples at $P(M) > 0.3$. The result is therefore not only robust across all thresholds but grows more statistically compelling as the sample expands, provided a meaningful M-type probability cut is maintained.
At $P(M) > 0$, the geometric mean of the 1855-object sample ($2.008\;\mathrm{d}^{-1}$) is indistinguishable from the null mean ($2.053\;\mathrm{d}^{-1}$), yielding $P = 0.78$. This most inclusive sample contains a large fraction of objects with only marginal M-type probability, and the signal is fully diluted. Taken together, these results confirm that the rotational excess is concentrated in high-confidence M-type candidates and erodes gradually and predictably as the sample is extended to lower-probability classifications.

\begin{table}[h]
\centering
\begin{tabular}{cccc}
\hline
$P(M)$ threshold & $n$ & GM (d$^{-1}$) & MC $p$-value \\
\hline
$>$0.8 &  76 & 2.576 & 0.044 \\
$>$0.7 & 117 & 2.736 & 0.005 \\
$>$0.6 & 137 & 2.762 & 0.003 \\
$>$0.5 & 172 & 2.713 & 0.001 \\
$>$0.4 & 194 & 2.712 & $<$0.001 \\
$>$0.3 & 240 & 2.673 & $<$0.001 \\
$>$0.2 & 276 & 2.602 & $<$0.001 \\
$>$0.1 & 332 & 2.414 & 0.01 \\
$>0$ & 1855 & 2.008 & 0.781 \\
\hline
\end{tabular}
\caption{Sensitivity of the MC result to the $P(M)$ classification 
threshold. As the threshold is lowered, the sample grows through the 
inclusion of progressively lower-confidence M-type candidates. The geometric mean rotation frequency decreases monotonically with the threshold, consistent with dilution. The non-monotonic behaviour of the p-value 
reflects the competing effect of increasing sample size, which narrows 
the null distribution and partially compensates for the loss of signal.}
\label{tab:purity}
\end{table}

\paragraph{Internal comparison} To further assess the robustness of the result against potential selection biases between heterogeneous catalogues, we perform an additional test using only the internal Mahlke sample. Instead of comparing M-type asteroids against the full SBDB population, we restrict both the metallic and non-M-type samples to objects included in the Mahlke catalogue. This ensures that all objects considered have undergone the same spectroscopic selection process and taxonomic classification, thereby minimising biases arising from differences in observational coverage, survey strategies, or data aggregation pipelines.

This approach, however, comes with a clear statistical caveat. While the resulting comparison is largely free from cross-catalogue observational biases, its conclusions are strictly limited to the subset of asteroids characterised in \cite{mahlke2022asteroid}. Extrapolating these results to the general asteroid population is therefore not straightforward. Accordingly, this analysis should be interpreted as a robustness check rather than a definitive test of the underlying hypothesis.

Within this internally consistent dataset, we repeat both Monte Carlo experiments described above: the diameter-conditioned test (MC) and the diameter-agnostic test (MC2). The results are summarised in Table \ref{tab:internal}.

\begin{table}[h]
\centering
\begin{tabular}{lcc}
\hline
 & MC (conditioned) & MC2 (agnostic) \\
\hline
$n_{\mathrm{M}}$            & 172   & 172   \\
$n_{\mathrm{nonM}}$         & 2439  & 2439  \\
GM (d$^{-1}$)               & 2.713 & 2.713 \\
Null mean                   & 2.038 & 2.115 \\
CI$_{2.5}$                  & 1.713 & 1.774 \\
CI$_{97.5}$                 & 2.397 & 2.485 \\
p-value                     & 0.0001 & 0.0011 \\
\hline
\end{tabular}
\caption{Monte Carlo results using only the internal Mahlke sample with a uniform quality filter ($U \geq 2$). GM: geometric mean rotation frequency of the M-type sample. MC: diameter-conditioned test. MC2: diameter-agnostic test. All tests use 10,000 iterations.}
\label{tab:internal}
\end{table}

The results not only reproduce but reinforce the signal obtained in Sections \ref{sec:MC} and \ref{sec:MC2}: both tests yield higher significance levels when the comparison is restricted to the non-M-type Mahlke population, with the conditioned test reaching $P=10^{-4}$, roughly an order of magnitude below the corresponding value obtained against the full SBDB background. In both cases, the geometric mean rotation frequency of the M-type sample remains well above the upper tail of the null distributions. This convergence across two independent background definitions provides evidence that the result is not driven by cross-catalogue inconsistencies or selection effects, but instead reflects an intrinsic difference between metallic and non-metallic asteroids.

\paragraph{Jackknife check}
As a final robustness check, we apply a leave-one-out jackknife procedure to assess whether the observed signal is driven by any single object in the metallic sample. In each of the 172 iterations, one M-type asteroid is removed and both MC and MC2 are re-run on the remaining 171-object subsample. For each iteration we record the geometric mean rotation frequency of the reduced sample and the corresponding one-sided p-values. If the result were dominated by an individual fast rotator, one would expect at least some iterations to yield substantially weaker significance levels; a narrow distribution of p-values across all leave-one-out subsamples would instead indicate that the excess rotation signal is a collective property of the metallic population rather than an artifact of any particular object.

\begin{figure}
    \centering
    \includegraphics{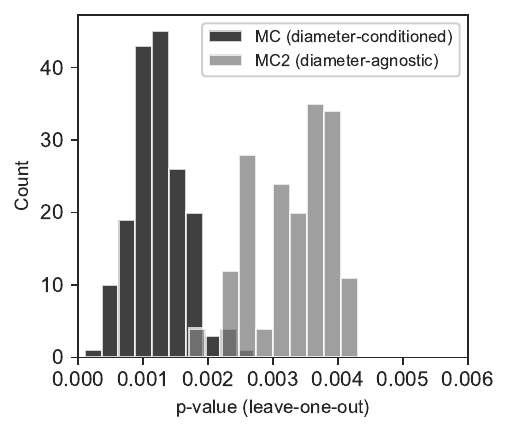}
    \caption{Distribution of one-sided p-values obtained from the leave-one-out jackknife procedure applied to the M-type sample ($n=172$). In each iteration, one M-type asteroid is removed and both the diameter-conditioned test (MC, dark) and the diameter-agnostic test (MC2, light) are re-run on the remaining 171-object subsample. All 172 iterations yield $P<0.01$ for both tests, with maximum p-values of 0.0027 (MC) and 0.0043 (MC2), confirming that the rotational excess of M-type asteroids is not driven by any individual object.}
    \label{fig:jackknife}
\end{figure}

The results are summarised in Table~\ref{tab:jackknife}. The geometric mean rotation frequency varies only between 2.692 and $2.768\;\textrm{d}^{-1}$ across all 172 subsamples, indicating that no single object exerts a disproportionate influence on the sample statistic. Both MC and MC2 p-values remain well below 0.01 in every iteration---100\% of leave-one-out runs satisfy this threshold---with maximum p-values of 0.0027 and 0.0043, respectively. The signal is therefore a collective property of the metallic sample and cannot be attributed to any individual asteroid. The corresponding p-value distributions are shown in Figure \ref{fig:jackknife}. 

\begin{table}[h]
\centering
\begin{tabular}{lcc}
\hline
 & MC (conditioned) & MC2 (agnostic) \\
\hline
$\bar{\omega}_g$ range (d$^{-1}$) & $[2.692,\ 2.768]$ & $[2.692,\ 2.768]$ \\
$P_{\min}$                             & $0.0001$           & $0.0017$           \\
$P_{\max}$                             & $0.0027$           & $0.0043$           \\
Fraction $P < 0.01$                    & $100\%$            & $100\%$            \\
\hline
\end{tabular}
\caption{Summary of the leave-one-out jackknife procedure applied to the
         M-type sample ($n = 172$). In each of the 172 iterations one object
         is removed and both Monte Carlo tests are re-run on the remaining
         subsample. Rows report the range of the geometric mean rotation
         frequency ($\bar{\omega}_g$) and of the one-sided p-values across
         all iterations, together with the fraction of runs yielding
         $p < 0.01$.}
\label{tab:jackknife}
\end{table}

\paragraph{Spin barrier proximity}
Although the purpose of this paper is to characterise the overall rotational behaviour of M-type asteroids rather than their fastest rotators, it is nevertheless worth examining, in a complementary manner, whether the Mahlke sample suggests that M-types approach their theoretical spin limit more closely. We examined the proximity of each taxonomic group in the Mahlke sample to its corresponding theoretical cohesionless spin barrier, which is classically derived from an equilibrium condition between gravity and centripetal force at the asteroid's equator, assuming an idealised spherical body with uniform density \citep{pravec2000fast}:

$$\omega_\textrm{crit}=\sqrt{\frac{4\pi G\rho}{3}}$$

\begin{figure}[h]
\centering
\includegraphics{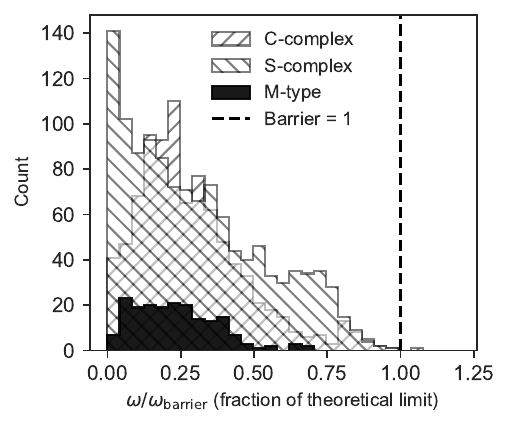}
\caption{Distribution of the rotation rate normalised to the theoretical cohesionless spin barrier, $\omega / \omega_{\mathrm{crit}}$, for the main taxonomic complexes in the Mahlke sample. Bulk densities for each taxonomic class have been obtained from \cite{carry2012density}. All groups are confined well below the disruption limit, with no objects approaching $\omega / \omega_{\mathrm{crit}} = 1$. M-type asteroids reach at most $\sim$0.7 of the theoretical barrier, with a 90th percentile of $\sim$0.4, indicating that the sample probes the sub-critical rotation regime.}
\label{fig:barrier_proximity}
\end{figure}

For each asteroid in the Mahlke dataset, we computed the ratio between its rotation frequency and the critical frequency expected for a strengthless body of the same bulk density, adopting representative values for the main taxonomic complexes. We then analysed the distribution of this normalised quantity across the sample, focusing on summary statistics such as the median, upper percentiles, and maximum values for each group. The results are shown in Figure \ref{fig:barrier_proximity} and listed in Table \ref{tab:proximity}.

\begin{table}
\centering
\begin{tabular}{lccc}
\hline
Taxonomic class & Median & 90th pct & Max \\
\hline
M-type        & 0.22 & 0.41 & 0.70 \\
S-complex     & 0.26 & 0.68 & 0.97 \\
C-complex     & 0.24 & 0.53 & 1.21 \\
\hline
\end{tabular}
\caption{Fraction of the theoretical cohesionless spin barrier reached ($\omega / \omega_{\rm crit}$) for the main taxonomic complexes. Values correspond to the median, 90th percentile, and maximum of each distribution. M-type asteroids do not appear to approach the theoretical limit more closely than other classes.}
\label{tab:proximity}
\end{table}

All taxonomic groups remain well below the theoretical disruption limit in their central behaviour, with median values around 0.2–0.3 of the spin barrier. While S- and C-complex asteroids extend to higher fractions, M-type asteroids remain confined to lower values, with a maximum of $\sim$0.7. No evidence is found in the Mahlke sample that M-type asteroids approach the spin barrier more closely than other taxonomic classes.

\paragraph{Family membership check} To assess whether the rotational excess could be driven by asteroid families — whose members share a common collisional origin and may inherit correlated spin properties — we cross-matched the M-type sample against the family assignments compiled in the LCDB, which adopts a hybrid of the \cite{nesvorny2015identification} and \cite{knezevic2012asteroids} dynamical family catalogues based on proper orbital elements. Objects assigned a Family Identification Number (FIN) below 9000 were classified as family members; those with FIN $\geq$ 9000 belong to broad orbital groups with no identified collisional parent and are treated as background objects. Of the 172 M-type candidates, 33 (19\%) are associated with a known dynamical family. Repeating both Monte Carlo tests on the remaining 139 family-free objects yields the results summarised in Table \ref{tab:family}. The p-values are essentially unchanged from those obtained with the full sample, confirming that the rotational excess is not driven by the collective contribution of any particular family and reflects an intrinsic property of the metallic population as a whole.
\begin{table}[h]
\centering

\begin{tabular}{lcccc}
\hline
Sample & $n$ & GM (d$^{-1}$) & MC $p$-val. & MC2 $p$-val. \\
\hline
Full sample     & 172 & 2.713 & 0.0014 & 0.0032 \\
Family-free     & 139 & 2.795 & 0.0013 & 0.0045 \\
\hline
\end{tabular}
\caption{Monte Carlo results for the full M-type sample and the
         family-free subsample. GM: geometric mean rotation frequency.
         MC: diameter-conditioned test. MC2: diameter-agnostic test.
         Family membership follows the LCDB hybrid catalogue.}
\label{tab:family}
\end{table}

\section{Discussion}
The statistical evidence presented in Section~\ref{sec:methods} points to a genuine rotational excess in the M-type population, one that persists across classification thresholds, reference samples, and methodological choices, and that cannot be attributed to family-specific effects or the contribution of any individual object. The central question is therefore not whether this excess exists, but what drives it, and whether the present dataset is sufficient to constrain its origin.

One aspect that the current sample is poorly suited to address is the high-frequency tail of the rotation distribution. No taxonomic group in the Mahlke catalogue approaches the theoretical disruption limit, and no clear differences are observed between M-type and non-M-type asteroids near the spin barrier. However, this result is not conclusive: the Mahlke catalogue is intrinsically biased toward larger objects, while the small, fast-rotating asteroids that probe the spin barrier are underrepresented due to the observational challenges of spectroscopic classification. The catalogue is therefore better suited to characterising the central behaviour of the rotation distribution than its high-frequency tail, and the present analysis cannot rule out the existence of M-type asteroids exceeding the typical spin limits observed in other taxonomic classes. Future large-scale surveys, such as the Legacy Survey of Space and Time (LSST) conducted by the Vera C. Rubin Observatory, are expected to provide a more complete census of small, fast-rotating asteroids and will be crucial to assess whether M-type bodies systematically approach or exceed the spin barrier.

\subsection{Statistical limitations}

\label{sec:limitations}

It must be acknowledged that the limited number of M-type asteroids known so far and the observational bias in their identification---with larger, brighter asteroids more likely to be identified in catalogues like Mahlke's---are constraining factors in the generalization of the faster spin of M-type asteroids. The lack of a representative M-type sample in lower diameter regimes is the weakest part of this analysis; we cannot exclude the possibility that the rotational excess reported here weakens or disappears entirely for asteroids 
with diameter $D < 2$ km. Also, it cannot be assumed that Mahlke's classification is perfect, and one should account for the possibility of a certain degree of misclassification.

Family-level structure also introduces an additional layer of complexity. Some M-types belong to collisional families or clusters whose members share dynamical histories and may inherit similar spin characteristics independent of composition. If specific families happen to contain an overabundance of fast rotators---whether due to collisional evolution, YORP-driven pathways, or stochastic events---they could contribute disproportionately to the observed signal. The potential influence of family membership is explicitly tested in Section \ref{sec:robustness}, where both Monte Carlo experiments are repeated on the 139 M-type asteroids with no known dynamical family association. The result is essentially unchanged: p-values of $P=0.0013$ (MC) and $P=0.0045$ (MC2), confirming that the rotational excess is not driven by the collective contribution of any particular family.

An additional source of bias may arise from the surface-dependent nature of taxonomic classification and its possible correlation with rotation rate. If faster rotating asteroids are more efficient at redistributing or partially removing regolith \citep{sanchez2020cohesive}, metallic-rich material could be more easily exposed at the surface. In this scenario, intrinsically metallic bodies with slower rotation rates might retain thicker mantling layers and therefore be less likely to be identified as M-types based on spectral and albedo criteria. Such a mechanism would preferentially enhance the detectability of fast-rotating metallic asteroids without necessarily reflecting a true absence of slowly rotating metal-rich bodies, effectively introducing a systematic bias in this analysis. While this effect cannot be quantified with the present dataset, it highlights the importance of distinguishing between surface expression and bulk composition when interpreting population-level trends.

\subsection{Physical implications}
The physical interpretation of this result is consistent with expectations about the internal structure of M-type bodies. First, higher material strength and cohesion in M-type asteroids should allow them to withstand higher rotation rates before reaching the disruption limit \citep{sanchez2020cohesive}. This contrasts with more porous or rubble-pile asteroids, whose rotation is limited by lower tensile strength. Second, collisional history may play a role: metallic objects could be more resistant to catastrophic fragmentation, allowing them to retain or even increase their spin after impacts that would otherwise disrupt less cohesive bodies. 

Although secular torques such as YORP can play an important role in asteroid spin evolution, their influence is strongly modulated by an object’s shape, surface properties, and internal density distribution. The YORP effect is most effective for bodies with high surface-to-mass ratios, low densities, and reflective surfaces, conditions that tend to amplify the thermal re-emission asymmetries responsible for spin-up \citep{bottke2006yarkovsky}.

In contrast, M-type asteroids commonly display higher densities, moderate albedos, and in some cases higher thermal conductivity \citep{opeil2010thermal},  all of which should reduce the magnitude of YORP accelerations. A more efficient redistribution of heat would further suppress the anisotropies required to sustain strong YORP torques, suggesting that YORP spin-up may be comparatively less efficient for metal-rich bodies than for silicate or carbon-rich ones. If this is the case, the enhanced spin rates of M-types would need to arise from other processes, such as differences in internal structure, cohesive strength, or collisional history. However, we caution that this reasoning remains qualitative: without dedicated modelling of YORP timescales for asteroids with M-type physical properties, the contribution of YORP to the observed rotational excess cannot be formally excluded.
 
It is also worth considering the potential role of ferrovolcanism, a hypothesis proposed to explain the surface properties of bodies such as (16) Psyche \citep{johnson2020ferrovolcanism}. During the migration of metallic melt toward the surface, conservation of angular momentum would act as a transient spin-down mechanism. Conversely, concentrating high-density material toward the exterior increases the moment of inertia relative to a homogeneous body of identical mass, potentially making the object more resistant to subsequent spin-state modifications imparted by collisions or external torques \citep{nichols-fleming2024moment}. Whether the net rotational effect of ferrovolcanism is an acceleration or a deceleration therefore depends on the balance between these two---and possibly other---competing processes, and remains an open question that cannot be answered without a more exhaustive quantitative analysis.

In any case, ferrovolcanism is expected to occur only during the early stages of differentiation and under restrictive compositional conditions tied to core solidification \citep{jorritsma2025constraints}. As such, it cannot account for a systematic population-wide trend in rotation rates, and should be regarded as a potential modulating factor for individual large bodies rather than a general explanation for the observed excess.

It is difficult to identify a single physical mechanism capable of explaining the excess rotation rates observed in M-type asteroids. Processes such as collisional evolution, YORP torques, or ferrovolcanism may influence individual objects, but they are either size-dependent, transient and without a clear, unilateral effect on rotation rate, and therefore unlikely to produce a systematic population-level bias. By contrast, a more monolithic internal structure associated with metal-rich compositions provides a simpler explanation: higher strength and cohesion allow metallic bodies to sustain faster rotation and to better preserve their spin states against disruption and external torques over long timescales. This interpretation, however, remains to be confirmed by complementary evidence from meteoritic analogs, radar sounding, 
or \textit{in situ} measurements.

\section{Summary and conclusions}
In this work we have investigated whether M-type asteroids exhibit systematically higher rotation frequencies than the general asteroid population. Using Mahlke’s probabilistic taxonomy in combination with diameter and rotation data from the SBDB and LCDB, we assembled the largest sample to date of candidate M-type asteroids with measured rotation rates. To account for the well-known size dependence of asteroid rotation and the observational size bias affecting the M-type sample, we applied a diameter-conditioned Monte Carlo resampling approach (MC), in which reference subsamples are drawn to match the diameter distribution of the M-type sample. The geometric mean rotation frequency of the M-type sample is $2.713\;\mathrm{d}^{-1}$, compared to a null distribution mean of $2.074\;\mathrm{d}^{-1}$, with a one-sided p-value of $P=0.0014$. A supplementary diameter-agnostic test (MC2) yields a consistent result ($P=0.0032$). The result is robust across a series of additional checks: it holds for all classification thresholds between $P(M)>0.2$ and $P(M)>0.8$; it is reproduced---and strengthened---when the comparison is restricted to the internal Mahlke sample, reaching $P=10^{-4}$; and a leave-one-out jackknife confirms that no individual asteroid drives the signal, with all 172 iterations yielding $P<0.01$. The convergence of these results across independent methodological choices and reference samples provides consistent statistical evidence that M-type asteroids rotate, on average, faster than the general asteroid population. The robustness checks presented in Section \ref{sec:robustness} make it unlikely that this excess is driven by sample selection, size bias, or classification threshold alone, though the limitations discussed in Section \ref{sec:limitations} preclude a definitive interpretation at this stage.

The interpretation of this result must, however, be tempered by the limitations of the current dataset. The known population of M-types remains small, their identification is subject to non-uniform observational biases, and taxonomic assignments are probabilistic rather than definitive. Differences in family membership, survey completeness, and period reliability also introduce sources of uncertainty that cannot be fully excluded. Nonetheless, the magnitude and consistency of the signal across multiple statistical frameworks indicate that these factors alone are unlikely to account for the observed excess in rotation rate.

Overall, our results provide the strongest statistical evidence to date that M-type features are linked to enhanced rotation rates. This connection may have implications for the internal strength of asteroids showing metallic features, the collisional and dynamical evolution of the asteroid belt, and the interpretation of forthcoming observations of (16) Psyche and other X-complex objects. Future work---including improved taxonomic coverage, more complete rotation surveys, and \textit{in situ} data taken by NASA's Psyche mission will help further clarify the origin and extent of the rotational properties identified here. In addition, the upcoming Legacy Survey of Space and Time (LSST) conducted by the Vera C. Rubin Observatory is expected to increase the known population of small Solar System bodies by more than an order of magnitude, discovering several million new asteroids and comets \citep{ivezic2019lsst}. This considerable expansion of sample size will provide a critical benchmark for assessing the robustness of the trends reported in this work.

\section*{Acknowledgements}
This work is based on research conducted as part of the author’s Master’s Thesis in Astronomy and Astrophysics at the Universitat Internacional Valenciana, supervised by Noemí Pinilla-Alonso and Miquel Serra-Ricart, and submitted on 30 October 2025. The author also acknowledges João Bento for their careful review of this work and for providing valuable feedback and suggestions.

The author acknowledges RICTEL TTT, S.A., owner of the TTT telescopes at the Teide Observatory, for its support during the course of this work. The company enabled the author to combine this research with graduate studies and provided financial support for the Master’s degree in the form of a loan.

This research has made use of the VizieR catalogue access tool, CDS, Strasbourg, France \citep{10.26093/cds/vizier}. The original description 
of the VizieR service was published in \citet{vizier2000}.

\appendix



\bibliographystyle{elsarticle-harv} 
\bibliography{example}






\end{document}